\documentclass[letterpaper]{article} % DO NOT CHANGE THIS
\usepackage{aaai24}  % DO NOT CHANGE THIS
\usepackage{times}  % DO NOT CHANGE THIS
\usepackage{helvet}  % DO NOT CHANGE THIS
\usepackage{courier}  % DO NOT CHANGE THIS
\usepackage[hyphens]{url}  % DO NOT CHANGE THIS
\usepackage{graphicx} % DO NOT CHANGE THIS
\usepackage{natbib}  % DO NOT CHANGE THIS AND DO NOT ADD ANY OPTIONS TO IT
\usepackage{caption} % DO NOT CHANGE THIS AND DO NOT ADD ANY OPTIONS TO IT
\usepackage{algorithm}
\usepackage{algorithmic}

\newtheorem{definition}{Definition}
\newcommand{\frameworkname}{MASTER} 
\usepackage{booktabs} 
\usepackage{bm}
\usepackage{multirow}
\usepackage{makecell}
\usepackage{newfloat}
\usepackage{listings}
\DeclareCaptionStyle{ruled}{labelfont=normalfont,labelsep=colon,strut=off} % DO NOT CHANGE THIS
\floatstyle{ruled}
\newfloat{listing}{tb}{lst}{}
\floatname{listing}{Listing}
\title{\frameworkname: Market-Guided Stock Transformer for Stock Price Forecasting}
\author{
    Tong Li\textsuperscript{\rm 1}\thanks{This work was done during her internship at Alibaba Group.},
    Zhaoyang Liu\textsuperscript{\rm 2},
    Yanyan Shen\textsuperscript{\rm 1}\thanks{Corresponding author.},
    Xue Wang\textsuperscript{\rm 2},
    Haokun Chen\textsuperscript{\rm 2},
    Sen Huang\textsuperscript{\rm 2}
}
\affiliations{
    \textsuperscript{\rm 1} Shanghai Jiao Tong University, 
    \textsuperscript{\rm 2} Alibaba Group \\
    \{2017lt, shenyy\}@sjtu.edu.cn,
    \{jingmu.lzy, xue.w, hankel.chk, huangsen.huang\}@alibaba-inc.com 
}

\usepackage{bibentry}
\begin{document}

\maketitle

\begin{abstract}
Stock price forecasting has remained an extremely challenging problem for many decades due to the high volatility of the stock market. 
Recent efforts have been devoted to modeling complex stock correlations toward joint stock price forecasting. Existing works share a common neural architecture that learns temporal patterns from individual stock series and then mixes up temporal representations to establish stock correlations. 
However, they only consider time-aligned stock correlations stemming from all the input stock features, which suffer from two limitations. First, stock correlations often occur momentarily and in a cross-time manner. Second, the feature effectiveness is dynamic with market variation, which affects both the stock sequential patterns and their correlations.
To address the limitations, this paper introduces
\frameworkname, a \textbf{MA}rkert-Guided \textbf{S}tock \textbf{T}ransform\textbf{ER}, which models the momentary and cross-time stock correlation and leverages market information for automatic feature selection. \frameworkname~elegantly tackles the complex stock correlation by alternatively engaging in intra-stock and inter-stock information aggregation. 
Experiments show the superiority of \frameworkname~compared with previous works and visualize the captured realistic stock correlation to provide valuable insights.
\end{abstract}

\section{Introduction}
Stock price forecasting, which utilizes historical data collected from the stock market to predict future trends, is a vital technique for profitable stock investment.
Unlike stationary time series that often exhibit regular patterns such as periodicity and steady trends, the dynamics in the stock price series are intricate because stock prices fluctuate subject to multiple factors, including macroeconomic factors, capital flows, investor sentiments, and events. 
The mixing of factors interweaves the stock market as a correlated network, making it difficult to precisely predict the individual behavior of stocks without taking other stocks into account. 

\begin{figure}
    \centering
    \includegraphics[width=\linewidth]{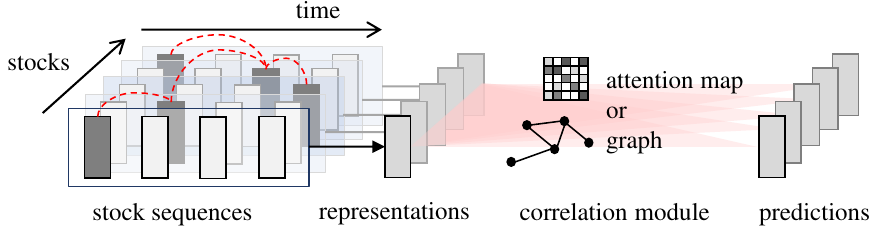}
    \caption{The framework of existing works. The dashed lines represent the underlying momentary and cross-time stock correlations, which reside between some \textit{($stock_1, time_1$)}, \textit{($stock_2, time_2$)} pairs.}
    \label{fig:ts} 
\end{figure}

Most previous works~\cite{feng2019temporal, xu2021hist, wang2021hierarchical, wang2022adaptive, wang2022review} in the field of stock correlation have relied on predefined concepts, relationships, or rules and established a \emph{static} correlation graph, e.g., stocks in the same industry are connected to each other.
While these methods provide insights into the relations between stocks, they do not account for the real-time correlation of stocks.
For example, different stocks within the same industry can experience opposite price movements on a particular day.
Additionally, the pre-defined relationships may not be generalizable to new stocks in an evolving market where events such as company listing, delisting, or changes in the main business happen normally.
Another line of research~\cite{yoo2021accurate} follows the Transformer architecture~\cite{vaswani2017attention}, and use the self-attention mechanism to compute \emph{dynamic} stock correlations. 
This data-driven manner is more flexible and applicable to the time-varying stock sets in the market.
Despite different schemes for establishing stock correlations, the existing methods generally follow a common two-step computation flow. As depicted in Figure~\ref{fig:ts}, the first step is using a sequential encoder to summarize the historical sequence of stock features, and obtain stock representation, and the second step is to refine each stock representation by aggregating information from correlated stocks using graph encoders or attention mechanism. 
However, such a flow suffers from two limitations.

%left bottom right top
%trim=280 210 200 150

First, existing works distill an overall stock representation and blur the time-specific details of stock sequence, leading to weakness in modeling the de-facto stock correlations, which often occurs \emph{momentarily} and in a \emph{cross-time} manner~\cite{bennett2022lead}.
To be specific, the stock correlation is highly dynamic and may reside in misaligned time steps rather than holding true through the whole lookback period.
This is because the dominating factors of stock prices constantly change, and different stocks may react to the same factors with different delays. For instance, upstream companies' stock prices may react faster to a shortage of raw materials than those of downstream companies, and individual stocks exhibit a lot of catch-up and fall-behind behaviors. 

Since the stock correlation may underlie between every stock pair and time pair, a straightforward way to simulate the momentary and cross-time correlation is to gather the $\tau \times |\mathcal{S}|$ feature vectors for pair-wise attention computation, where $\tau$ is the lookback window length and $\mathcal{S}$ is the stock set.
However, in addition to the increased computational complexity, this approach faces practical difficulties because the stock forecasting task is in intense data hunger. 
Intuitively, there are only around 250 trading days per year, producing limited observations on stocks.
When the model adopts such a large attention field with insufficient training samples, it often struggles to optimize and may even fall into suboptimal solutions. 
Although clustering approaches like local sensitive hashing~\cite{kitaev2020reformer} have been proposed to reduce the size of the attention field, they are sensitive to initialization, which is a fatal issue in a data-hungry domain like stock forecasting.
To address these challenges, we propose a novel stock transformer architecture specifically designed for stock price forecasting. Rather than directly modeling the $\tau \times |\mathcal{S}|$ attention field or using clustering-based approximation methods, our model aggregates information from different time steps and different stocks alternately to model realistic stock correlation and facilitate model learning.

Another limitation of existing works is that they ignore the impact of varying market status.
In long-term practice with the market variation, one essential observation by investors is that the features come into effect and expire dynamically. 
The effectiveness of features has an influence on both the intra-stock sequential pattern and the stock correlation. 
For instance, in a bull market, the correlations among stocks are more significant due to the investors' optimism.
Traditional investors repeatedly conduct statistical examination on to select effective feature, which is exhaustive and face a gap when integrated with learning-based methods. 
To save the human efforts, we are motivated to equip our stock transformer with a novel gating mechanism, which incorporates the market information to perform automatically feature selection.  
We name the proposed method \frameworkname, standing for \textbf{MA}rket-Guided \textbf{S}tock \textbf{T}ransform\textbf{ER}. 
To summarize, our main contributions are as follows.

$\bullet$ We propose a novel stock transformer for stock price forecasting to effectively capture the stock correlation. To the best of our knowledge, we are the first to mine the momentary and cross-time stock correlation with learning-based methods. 

$\bullet$ We introduce a novel gating mechanism that integrates market information to automatically select relevant features and adapt to varying market scenarios. 

$\bullet$ We conducted experiments to validate the designs of our proposed method and demonstrated its superiority compared to baselines. The visualization results provided valuable insights into the real-time dynamics of stock correlations.
 
\begin{figure*}
    \centering
    \includegraphics[width=\textwidth]{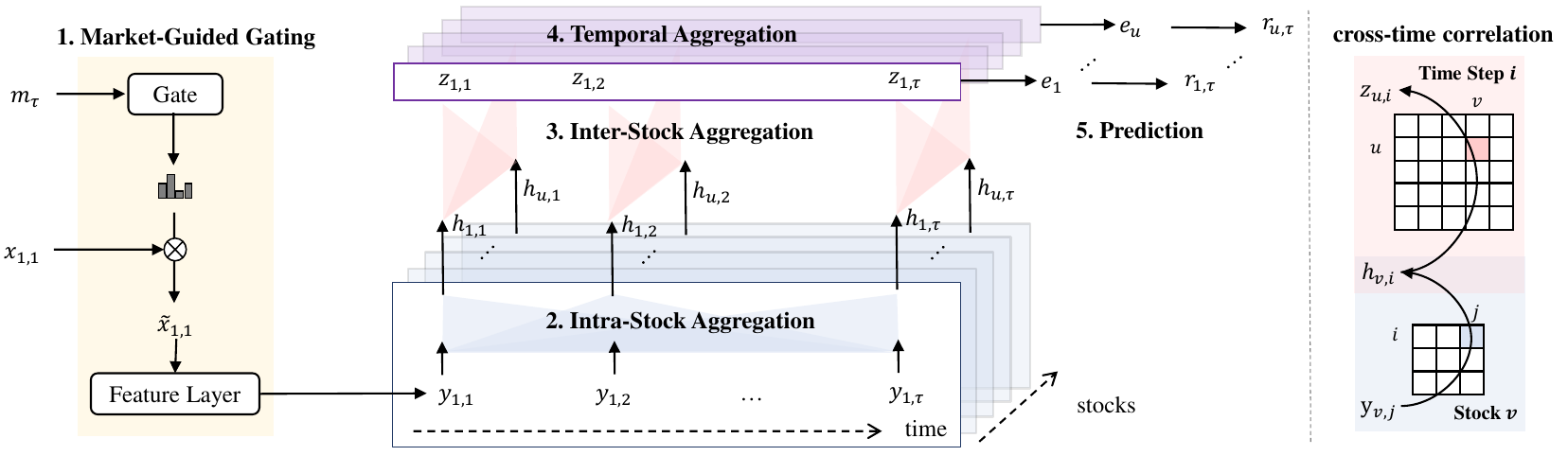}
    \caption{Overview of the \frameworkname~framework.}
    \label{fig:framework}
\end{figure*}
\section{Methodology}

\subsection{Problem Formulation}
The indicators of each stock $u\in \mathcal{S}$ are collected at every time step $t\in[1,\tau]$ to form the feature vector $x_{u,t} \in \mathbf{R}^F$.
Following existing works on stock market analysis \cite{feng2018enhancing, sawhney2020spatiotemporal, huynh2023efficient}, we focus on the prediction of the change in stock price rather than the absolute value. 
The return ratio, which is the relative close price change in $d$ days, is $\tilde{r}_{u}=(c_{u,\tau+d}-c_{u,\tau+1})/c_{u,\tau+1}$, where $c_{u,t}$ is the closing price of stock $u$ at time step $t$, and $d$ represents the predetermined prediction interval.
%in application scenarios.
The return ratio normalizes the market price variety between different stocks in comparison to the absolute price change.
Since stock investment is to rank and select the most profitable stocks, we perform daily Z-score normalization of return ratio to encode the label with the rankings, $r_{u}={\rm Norm}_\mathcal{S}(\tilde{r}_{u})$, as in previous work~\cite{yang2020qlib}.

\begin{definition}[Stock Price Forecasting]
Given stock features $\{x_{u,t}\}_{u\in \mathcal{S},t\in [1, \tau]}$, the stock price forecasting is to jointly predict the future normalized return ratio $\{r_{u}\}_{u\in \mathcal{S}}$.
\end{definition}

\subsection{Overview}
Figure~\ref{fig:framework} depicts the architecture of our proposed method \frameworkname, which consists of five steps.
(1) \textbf{Market-Guided Gating}. We construct a vector representing the current market status $m_\tau$ and leverage it to rescale feature vectors by a gating mechanism, achieving market-guided feature selection.
(2) \textbf{Intra-Stock Aggregation}.
Within the sequence of each stock, at each time step, we aggregate information from other time steps to generate a local embedding that preserves the temporal local details of the stock while collecting all important signals along the time axis.
The local embedding $h_{u,t}$ will serve as relays and transport the collected signals to other stocks in subsequent modules.
(3) \textbf{Inter-Stock Aggregation}.
At each time step, we compute stock correlation with attention mechanism, and each stock further aggregates the local embedding of other stocks. 
The aggregated information $z_{u,t}$, which we refer to as temporal embedding, contains not only the information of the momentarily correlated stocks at $t$, but also preserves the personal information of $u$.
(4) \textbf{Temporal Aggregation}.
For each stock, the last temporal embedding queries from all historical temporal embedding and produce a comprehensive stock embedding $e_{u}$.
(5) \textbf{Prediction}.
The comprehensive stock embedding is sent to prediction layers for label prediction.
We elaborate on the details of \frameworkname~step by step in the following sub-sections.

\subsection{Market-Guided Gating}
\subsubsection{Market Status Representation}
First, we propose to combine information from two aspects into a vector $m_\tau$ to give an abundant description of the current market status.
(1) Market index price.
The market index price is a weighted average of the prices of a group of stocks $\mathcal{S'}$ by their share of market capitalization. 
$\mathcal{S'}$ is typically composed of top companies with the most market capitalization, representing a particular market or sector, and may differ from user-interested stocks in investing $\mathcal{S}$.
We include both the current market index price at $\tau$ and the historical market index prices, which is described by the average and standard deviation in the past $d'$ days to reveal the price fluctuations. Here, $d'$ specifies the referable interval length to introduce historical market information in applications. 
(2) Market index trading volume. The trading volumes of $\mathcal{S'}$ reveals the investors involvement, reflecting the activity of the market. We include the average and standard deviation of market index trading volume in the past $d'$ days, to reveal the actual size of the market. $\mathcal{S'}$ and $d'$ are identical to the aforementioned definitions.
Now we present the market-guided stock price forecasting task.

\begin{definition}[Market-Guided Stock Price Forecasting]
Given $\{x_{u,t}\}_{u\in \mathcal{S},t\in [1, \tau]}$ and the constructed market status vector $m_\tau$,  market-guided stock price forecasting is to jointly predict the future normalized return ratio $\{r_u\}_{u\in \mathcal{S}}$.
\end{definition}

\subsubsection{Gating Mechanism}
The gating mechanism generates one scaling coefficient for each feature dimension to enlarge or shrink the magnitude of the feature, thereby emphasizing or diminishing the amount of information from the feature flowing to the subsequent modules. 
The gating mechanism is learned by the model training, and the coefficient is optimized by how much the feature contributes to improve forecasting performance, thus reflect the feature effectiveness. 

Given the market status representation $m_\tau, |m_\tau|=F'$,  
we first use a single linear layer to transform $m_\tau$ into the feature dimension $F=|x_{u,t}|$.
Then, we perform Softmax along the feature dimension to obtain a distribution. 
$$ \alpha(m_\tau) = F\cdot {\rm softmax}_{\beta}(W_{\alpha}m_\tau+b_{\alpha}),$$
where $W_{\alpha}$, $b_{\alpha}$ are learnable matrix and bias, $\beta$ is the temperature hyperparameter controlling the sharpness of the output distribution.
Softmax compels a competition among features to distinguish the effective ones and ineffective ones.
Here, a smaller temperature $\beta$ encourages the distribution to focus on certain dimension and the gating effect is stronger while a larger $\beta$ makes the distribution incline to even and the gating effect is weaker. 
Note that we enlarge the value at each dimension by $F$ times as the scaling coefficient. This operation compare the generated distribution with a uniform distribution where each dimension is $1/F$, to determine whether to enlarge or shrink the value.  
The intuition to generate coefficients from $m_\tau$ is that the effectiveness of features are influenced by market status. For example, if the model learns moving average (MA) factor is useful during volatile market periods, it will emphasize MA when the market becomes volatile again.
Under the same $m_\tau$, $\alpha$ are shared for $\{x_{u,t}\}$, $u\in \mathcal{S}$, $t\in [1, \tau]$, in that we incorporate the most recent market status to perform unified feature selection. The rescaled feature vectors are $ \tilde{x}_{u,t} = \alpha(m_\tau) \circ x_{u,t}$, where $\circ$ is the Hadamard product. 

\subsection{Intra-Stock Aggregation}
In MASTER, we use intra-stock aggregation followed by inter-stock aggregation to break down the large and complex attention field.
Although the entire market is complicated with diverse behaviours of individual stocks, the patterns of a specific stock tend to be relatively continuous.
Therefore, we perform intra-stock aggregation first due to its smaller attention field and simpler distribution.
In our proposed intra-stock aggregation, the feature at each time step aggregate information from other time steps and form a local embedding. 
Compared with existing works which initially mix the feature sequence into one representation~\cite{yoo2021accurate}, we maintain a sequence of local embedding which are advised with the important signals in sequence through intra-stock aggregation while reserve the local details.

We first send the rescaled feature vectors to a feature encoder and transform them into the embedding space, $y_{u,t}=f(\tilde{x}_{u,t})$, $|y_{u,t}|=D$. We simply use a single linear layer as $f(\cdot)$.
Then, we apply a bi-directional sequential encoder to obtain the local output at each time step $t$.
Inspired by the success of transformer-based models in modeling sequential patterns, we instantiate the sequential encoder with a single-layer transformer encoder~\cite{vaswani2017attention}.
Each feature vector at a particular time step is treated as a token, and we add a fixed $D$-dimensional sinusoidal positional encoding $p_{t}$ to mark the chronically order in the look back window.  
$$Y_u=||_{t\in[1,\tau]}\textsf{LN}(f(\tilde{x}_{u,t})+p_t),$$ where $||$ denotes the concatenation of vectors and \textsf{LN} the layer normalization. Then, the feature embedding at each time step queries from all time steps in the stock sequence. 
We introduce multi-head attention mechanisms, denoted as \textsf{MHA}$(\cdot)$,  with $N_1$ heads to perform different aggregations in parallel. We also utilize feed forward layers, \textsf{FFN}$(\cdot )$, to fuse the information obtained from the multi-head attention.
$$ Q^1_u=W^1_QY_u,\quad K^1_u=W^1_KY_u,\quad V^1_u=W^1_VY_u,$$
$$ H^1_u=||_{t\in[1,\tau ]} h_{u,t} =\textsf{FFN}^1(\textsf{MHA}^1(Q^1_u,K^1_u,V^1_u)+Y_u),$$
where \textsf{FFN} is a two-layer MLP with ReLU activation and residual connection. As a result, the local embedding $h_{u,t}$ both reserve the local details and encode indicative signals from other time steps. 
%which will serve as a mediate to broadcast the signals to o. 

\subsection{Inter-Stock Aggregation}
Then, we consider aggregating information from correlated stocks. 
Compared with existing works that distill an overall stock correlation, we establish a series of momentary stock correlation corresponding to every time step.
Instead of using pre-defined relationships that face a mismatch with the proximity of real-time stock movements, we propose to mine the asymmetric and dynamic inter-stock correlation via attention-mechanism.
The quality of the correlation will be measured by its contribution to improving the forecasting performance, and automatically optimized by the model training process.

Specifically, at each time step, we gather the local embedding of all stocks $H^2_t=||_{u\in\mathcal{S}} h_{u,t}$ and perform multi-head attention mechanism with $N_2$ heads. 
$$Q^2_t = W^2_QH^2_t,\quad K^2_t=W^2_KH^2_t, \quad V^2_t=W^2_VH^2_t,$$
$$ Z_t=||_{u\in\mathcal{S}} z_{u,t} =\textsf{FFN}^2(\textsf{MHA}^2(Q^2_t,K^2_t,V^2_t)+H^2_t).$$
With the residual connection of \textsf{FFN}, the temporal embedding $z_{u,t}$ is encoded with both the information from momentarily correlated stocks and the personal information of stock $u$ itself. 
Our stock transformer is able to model the cross-time correlation of stocks, as shown in Figure~\ref{fig:framework} (Right). The local details of $y_{v,j}$ can first be conveyed to $h_{v,i}$ by the intra-stock aggregation of stock $v$, and then transmitted to $z_{u,i}$ by inter-stock aggregation at time step $i$, hence modeling the correlation from any $(v,j)$ to $(u,i)$. 
We further visualize and explain the captured cross-time correlation in the experiments section.

\subsection{Temporal Aggregation}
In contrast with existing works which obtain one embedding for each stock after modeling stock correlation~\cite{feng2019temporal}, our approach involves producing a series of temporal embedding $z_{u,t}, t\in[1,\tau].$ 
Each $z_{u,t}$ is encoded with information from stocks that are momentarily correlated with $(u,t)$.
To summarize the obtained temporal embeddings and obtain a comprehensive stock embedding $e_u$, we employ a temporal attention layer along the time axis. 
We use the latest temporal embedding $z_{u,\tau}$ as the query vector, and compute the attention score $\lambda_{u,t}$ in a hidden space with transformation matrix $W_{\lambda}$,
$$\lambda_{u,t}=\frac{\exp(z^T_{u,t}W_{\lambda}z_{u,\tau})}{\sum_{i\in [1,\tau]} \exp(z^T_{u,i}W_{\lambda}z_{u,\tau})},
\quad e_u=\sum_{t\in[1,\tau]}\lambda_{u,t} z_{u,t}.$$

\subsection{Prediction and Training}
Finally, the stock embedding $e_u$ is fed into a predictor $g(\cdot)$ for label regression. We use a single linear layer as the predictor, and the forecasting quality is measured by the MSE loss. 
In each batch, MASTER is jointly optimized for all $ u\in \mathcal{S}$ on a particular prediction date. And a training epoch is composed of multiple batches correspond to different prediction dates in the training set.
$$\hat{r}_u=g(e_u), \quad L=\sum_{u\in \mathcal{S}}\textsf{MSE}(r_u, \hat{r}_u).$$

\subsection{Discussions}
\subsubsection{Relationships with Existing Works}
%stock price forecasting
Modeling stock correlations has long been an indispensable research direction for stock price prediction. Today, many researchers and quantitative analysts, still opt for linear models, support vector machines, and tree-based methods for stock price forecasting~\cite{nugroho2014decision,chen2016xgboost,kamble2017short,xie2013semantic,li2015tensor, piccolo1990distance}. The aggregation of correlation information within and between stocks is often achieved through feature engineering, which relies heavily on manual expertise and constantly faces the risk of factor decay.
Inspired by the success of neural sequential data analysis, researchers are driven to take into account the stock feature sequences and learn the temporal correlation automatically. 
They design various sequential models, such as RNN-based \cite{feng2019temporal, sawhney2021stock, yoo2021accurate, huynh2023efficient}, CNN-based \cite{wang2021hierarchical}, and attention-based models\cite{liu2019transformer, ding2020hierarchical}, to mine the internal temporal dynamics of a stock. 
Recent research focus on the modeling of stock correlation, which add a correlation module in posterior to the sequential model as illustrated in Figure~\ref{fig:ts}.
They propose to use graph-based \cite{feng2019temporal, xu2021hist,wang2021hierarchical, wang2022adaptive}, hypergraph-based \cite{sawhney2021stock,huynh2023efficient} and attention-based \cite{yoo2021accurate, xiang2022temporal} modules to build the overall stock correlation and perform joint prediction. 
Our \frameworkname~is dedicated to momentary and cross-time stock correlation mining. To do so, we develop a novel model architecture as in Figure~\ref{fig:framework} that is genuinely different from all existing methods. 
Furthermore, \frameworkname~is specialized for stock price forecasting, which is distinct in data form and task properties from existing transformer-based models in spatial-temporal data ~\cite{bulat2021space, cong2021spatial, xu2020spatial,li2023memory} or multivariate time series domains~\cite{zhang2022crossformer, nie2022time}.

\subsubsection{Complexity Analysis}
We now analyze the computation complexity of our proposed method.
Let $M=|\mathcal{S}|$, the market-guided gating rescale $M \times \tau$ feature vectors of dimension $F$.
In intra-stock aggregation, the calculation amount of pair-wise attention is $\tau^2$ for each stock at each attention head.
In inter-stock aggregation, the calculation amount is $M^2$ at each time step and each attention head.
In temporal aggregation, we compute $\tau$ attention scores for each stock. 
The overall computation complexity is {\rm O}$(FM\tau+N_1M\tau^2D^2+N_2M^2\tau D^2+M\tau D^2)$, where $M\gg \tau$.
Therefore, \frameworkname~is of ${\rm O}(N_2M^2\tau D^2)$ time complexity.
Compared with directly operating on the $M \times \tau$ attention field with $N$ attention heads, which is in {\rm O}$(NM^2\tau^2D^2)$, we reduce the computation cost by about $\tau$ times and achieve modeling cross-time correlations between stocks more efficiently.
The overall parameters to be trained in \frameworkname~are transformation matrices $W^1_Q, W^1_K, W^1_V, W^2_Q, W^2_K, W^2_V, W_\lambda$, which is in shape $D\times D$, and parameters in MLP layers $\alpha, f, \textsf{FFN}^1, \textsf{FFN}^2$ and $g$.

\section{Experiments}
In this section, we conduct experiments to answer the following four research questions:
\begin{itemize}
    \item \textbf{RQ1} How is the overall performance of \frameworkname~compared with state-of-the-art methods?
    \item \textbf{RQ2} Is the proposed stock transformer architecture effective for stock price forecasting?
    \item \textbf{RQ3} How do hyper-parameter configurations affect the performance of \frameworkname?
    \item \textbf{RQ4} What insights on the stock correlation can we get through visualizing the attention map?
\end{itemize}

\subsubsection{Datasets}
We evaluate our framework on the Chinese stock market with CSI300 and CSI800 stock sets. CSI300 and CSI800 are two stock sets containing 300 and 800 stocks with the highest capital value on the Shanghai Stock Exchange and the Shenzhen Stock Exchange. 
The dataset contains daily information ranging from 2008 to 2022 of CSI300 and CSI800. 
We use the data from Q1 2008 to Q1 2020 as the training set, data in Q2 2020 as the validation set, and the last ten quarters, i.e., Q3 2020 to Q4 2022, are reserved as the test set.  We apply the public Alpha158 indicators \cite{yang2020qlib} to extract stock features from the collected data. 
The lookback window length $\tau$ and prediction interval $d$ are set as $8$ and $5$ respectively. 
For market representation, we constructed $63$ features with CSI300, CSI500 and CSI800 market indices, and  refereable interval $d'=5,10,20,30,60$.  

\subsubsection{Baselines}
We compare the performance of MASTER with several stock price forecasting baselines from different categories.
$\bullet$ XGBoost~\cite{chen2016xgboost}: A decision-tree based method. According to the leaderboard of Qlib platform~\cite{yang2020qlib}, it is one of the strongest baselines.
$\bullet$ LSTM~\cite{graves2012long}, GRU~\cite{cho2014learning}, TCN~\cite{bai2018empirical}, and Transformer~\cite{vaswani2017attention}: Sequential baselines that leverage vanilla LSTM/GRU/Temporal convolutional network/Transformer along the time axis for stock price forecasting. 
$\bullet$ GAT~\cite{velivckovic2017graph}: A graph-based baseline, which first use sequential encoder to gain stock presentation and then aggregate information by graph attention networks{\footnote{More discussion is provided in the supplementary materials.}}. 
$\bullet$ DTML~\cite{yoo2021accurate}: A state-of-the-art stock correlation mining method, which follows the framework in Figure~\ref{fig:ts}. DTML adopts the attention-mechanism to mine the dynamic correlation among stocks and also incorporates the market information into the modeling.

\subsubsection{Evaluation}
We adopt both ranking metrics and portfolio-based metrics to give a thorough evaluation of the model performance. Four ranking metrics, Information Coefficient (IC), Rank Information Coefficient (RankIC), Information Ratio based IC (ICIR) and Information Ratio based RankIC (RankICIR) are considered. IC and RankIC are the Pearson coefficient and Spearman coefficient averaged at a daily frequency. ICIR and RankICIR are normalized metrics of IC and RankIC by dividing the standard deviation. Those metrics are commonly used in literature (e.g., \citealt{xu2021hist} and \citealt{yang2020qlib}) to describe the performance of the forecasting results from the value and rank perspectives.
Furthermore, we employ two portfolio-based metrics to compare the investment profit and risk of each method. 
We simulate daily trading using a simple strategy that selects the top 30 stocks with the highest return ratio and reports the Excess Annualized Return (AR) and Information Ratio (IR) metrics.
AR measures the annual expected excess return generated by the investment, while IR measures the risk-adjusted performance of an investment.

\begin{table*}[t]
\centering

\small
\begin{tabular}{c|l|cccc|cc}
\toprule
 Dataset & Model & IC & ICIR & RankIC & RankICIR & AR & IR\\
 \midrule
\multirow{8}*{CSI300}& XGBoost &  $0.051\pm0.001$  & $0.37\pm0.01$ & $0.050\pm 0.001$ & $0.36\pm0.01$ & \underline{$0.23\pm0.03$} & $1.9\pm0.3$ \\

& LSTM & $0.049 \pm 0.001$ & \underline{$0.41\pm 0.01$} & $0.051\pm 0.002$ &  $0.41\pm 0.03$ & $0.20\pm 0.04$ & \underline{$2.0\pm 0.4$} \\
& GRU & $0.052\pm 0.004$ & $0.35\pm 0.04$ & $0.052\pm 0.005$ & $0.34\pm 0.04$ & $0.19\pm 0.04$ & $1.5 \pm 0.3$\\
& TCN & $0.050\pm 0.002$ & $0.33\pm 0.04$ & $0.049\pm 0.002$ & $0.31\pm 0.04$ & $0.18\pm 0.05$ & $1.4 \pm 0.5$ \\
& Transformer & $0.047 \pm 0.007$ & $0.39\pm 0.04$ & $0.051\pm 0.002$ & \underline{$0.42\pm 0.04$} & $0.22\pm 0.06$ & $2.0\pm 0.4$ \\
& GAT & \underline{$0.054\pm 0.002$} & $0.36\pm 0.02$ & $0.041\pm 0.002$ & $0.25\pm 0.02$ & $0.19\pm 0.03$ & $1.3 \pm 0.3$ \\
& DTML & $0.049\pm0.006$  &$0.33\pm0.04$ &  \underline{$0.052\pm 0.005$} & $0.33\pm0.04$ & $0.21\pm0.03$ & $1.7\pm0.3$ \\
& \frameworkname~ & \bm{$0.064^*\pm0.006$}  &\bm{$0.42\pm0.04$} &  \bm{$0.076^*\pm 0.005$} & \bm{$0.49\pm0.04$} & \bm{$0.27\pm0.05$} & \bm{$2.4\pm0.4$} \\ 
\midrule
\multirow{8}*{CSI800} & XGBoost & $0.040 \pm 0.000$ & $0.37\pm 0.01$ & $0.047\pm 0.000$ & $0.42\pm 0.01$ & $0.08\pm 0.02$ & $0.6\pm 0.2$\\
 & LSTM & $0.028 \pm 0.002$ & $0.32\pm 0.02$ & $0.039\pm 0.002$ & $0.41\pm 0.03$ & $0.09\pm 0.02$ & $0.9\pm 0.2$\\
 & GRU & $0.039\pm 0.002$ & $0.36\pm 0.05$ & $0.044\pm 0.003$ & $0.39\pm 0.07$ & $0.07\pm 0.04$ & $0.6 \pm 0.3$\\
& TCN & $0.038\pm 0.002$ & $0.33\pm 0.04$ & $0.045\pm 0.002$ & $0.38\pm 0.05$ & $0.05\pm 0.04$ & $0.4 \pm 0.3$ \\
   & Transformer & $0.040 \pm 0.003$ & \bm{$0.43\pm 0.03$} & $0.048\pm 0.003$ & \bm{$0.51\pm 0.05$} & $0.13\pm 0.04$&$1.1\pm 0.3$\\
& GAT & \underline{$0.043\pm 0.002$} & $0.39\pm 0.02$ & $0.042\pm 0.002$ & $0.35\pm 0.02$ & $0.10\pm 0.04$ & $0.7 \pm 0.3$ \\
   & DTML & $0.039 \pm 0.004$ & $0.29\pm 0.03$ & \underline{$0.053\pm 0.008$} & $0.37\pm 0.06$ & \underline{$0.16\pm 0.03$} & \underline{$1.3\pm 0.2$}\\
& \frameworkname~ & \bm{$0.052^*\pm0.006$}  &\underline{$0.40\pm0.06$} &  \bm{$0.066\pm 0.007$} & \underline{$0.48\pm0.06$} & \bm{$0.28^*\pm0.02$} & \bm{$2.3^*\pm0.3$} \\ 
\bottomrule
\end{tabular}
\caption{Overall performance comparison. The best results are in bold and the second-best results are underlined. And * denotes statistically significant improvement (measured by t-test with p-value $<$ 0.01) over all baselines.}
\label{tab:performance}
\end{table*}

\begin{table*}[t]
\centering
%\small
\begin{tabular}{l|cccc|cc}
\toprule
 Model & IC & ICIR & RankIC & RankICIR & AR & IR\\
\midrule
(MA)STER  & \bm{$0.064\pm0.003$}  &\bm{$0.43\pm0.02$} &  \bm{$0.074\pm 0.004$} & \bm{$0.48\pm0.04$} & \bm{$0.25\pm0.03$} & \bm{$2.1\pm0.3$}\\
(MA)STER-Bi &\underline{$0.058\pm0.005$} & \underline{$0.38 \pm 0.04$} &\underline{$0.066\pm 0.008$} & \underline{$0.41\pm 0.05$} &\underline{$0.19\pm 0.03$} & $1.6\pm 0.2$\\
Naive &  $0.041\pm0.008$ &$0.30\pm0.05$ & $0.046\pm0.007$ & $0.32\pm0.04$ & $0.18\pm0.05$ & $1.6\pm0.6$\\
Clustering & $0.044\pm 0.003$ & $0.36 \pm 0.02$ & $0.049\pm 0.005$ & $0.39\pm 0.04$ & $0.18\pm 0.04$ & \underline{$1.7 \pm 0.3$}\\
\bottomrule
\end{tabular}
\caption{Experiments on CSI300 to validate the effectiveness of proposed stock transformer architecture. The best results are in bold and the second-best results are underlined.}
\label{tab:effect}
\end{table*}

\subsubsection{Implementation}
We implemented \frameworkname{\footnote{Code and supplementary materials are at \url{https://github.com/SJTU-Quant/MASTER}}}~with PyTorch and build our methods based on the open-source quantitative investment platform Qlib~\cite{yang2020qlib}. For DTML, we implement it based on the original paper since there is no official implementation publicly. For other baselines, we use their Qlib implementations.
For hyperparameters of each baseline method, the layer number and model size are tuned from $\{1,2,3\}$ and $\{128, 256, 512\}$ respectively. The learning rate $lr$ is tuned among $\{10^{-i}\}_{i\in\{3,4,5,6\}}$, and we selected the best hyperparameters based on the IC performance in the validation stage.
For hyperparameters of \frameworkname, we tune the model size $D$ and learning rate $lr$ among the same range as the baselines, and the final selection is $D$=$256$, $lr$=$10^{-5}$ for all datasets; we set $N_1$=$4$, $N_2$=$2$ for all datasets and $\beta$=$5$ and $\beta$=$2$ for CSI300 and CSI800 respectively. More implementation details of baseline methods are summarized in the supplementary materials.
Each model is trained for at most $40$ epochs with early stopping.
All the experiments are conducted on a server equipped with Intel(R) Xeon(R) Platinum 8163 CPU, 128GB Memory, and a Tesla V100-SXM2 GPU (16GB Memory). 
Each experiment was repeated 5 times with random initialization and the average performance was reported.

\subsection{Overall Performance (RQ1)}
\label{sec:interval}
The overall performance are reported in Table~\ref{tab:performance} %shows the overall performance of compared methods.\label{tab:performance}
\frameworkname~achieves the best results on 6/8 of the ranking metrics, and 
consistently outperforms all benchmarks in the portfolio-based metrics. In particular, \frameworkname~achieve $13\%$ improvements in ranking metrics and $47\%$ improvements in portfolio-based metrics compared to the second-best results on the average sense. Note that ranking matrics are computed with the whole set and portfolio-based metrics mostly consider the 30 top-performed stocks. 
The achievements in both types of metrics imply that MASTER is of good predicting ability on the whole stock set without sacrificing the accuracy of the important stocks. 
The significant improvements cast light on the importance of stock correlation modeling, so each stock can also benefit from the historical signals of other momentarily correlated stocks. 
We also observe all methods gain better performance on CSI300 over CSI800. We believe it is because CSI300 consists of companies with larger capitalization whose stock prices are more predictable. 
When compared to the existing stock correlation method (i.e., DTML), \frameworkname~outperforms in all 6 metrics, which tells our proposed Market-Guided Gating and aggregation techniques are more efficient in mining cross-stock information than existing literature. 

%left bottom right top
\begin{figure*}[]
    \includegraphics[width=\linewidth]{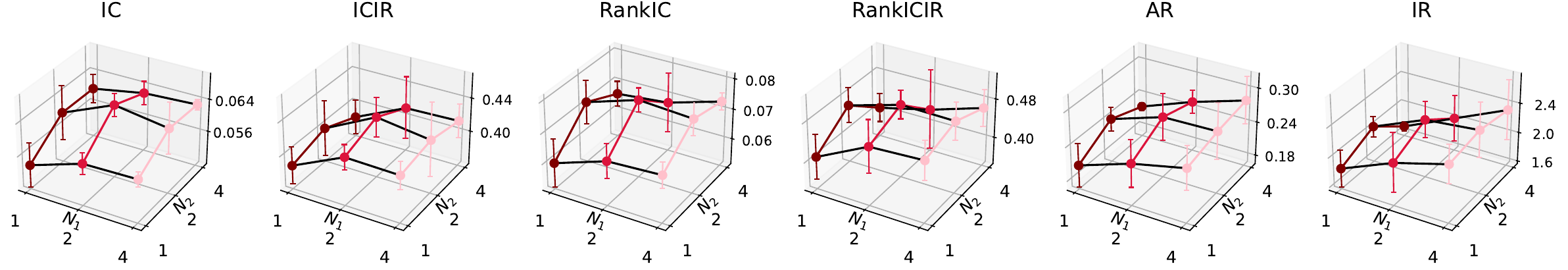}
\caption{The average and standard deviation of metrics with different $(N_1,N_2)$ combinations on CSI300.}
\label{fig:mh}
\end{figure*}

\begin{figure*}[]
    \includegraphics[width=\linewidth]{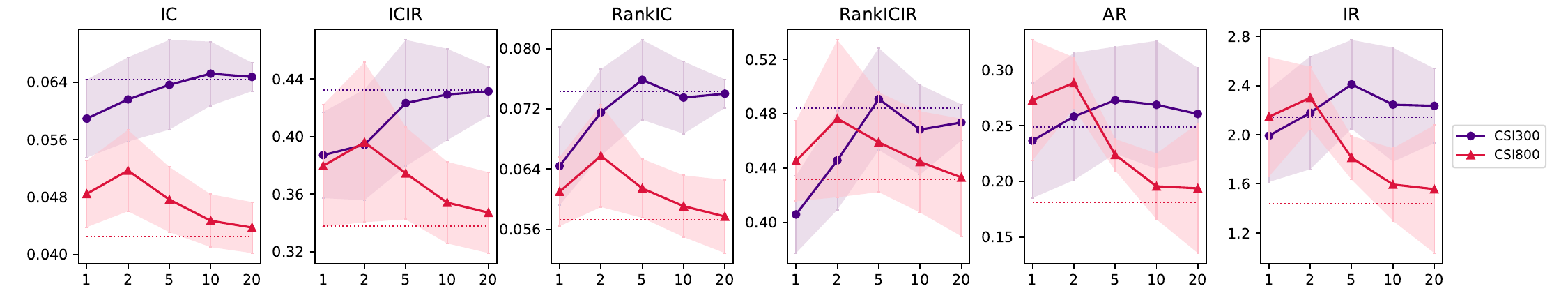}
\caption{\frameworkname~performance with varying $\beta$. The horizontal dash lines are performance without market-guided gating.}
\label{fig:beta}
\end{figure*}

\begin{figure}[t]
    \includegraphics[width=\linewidth]{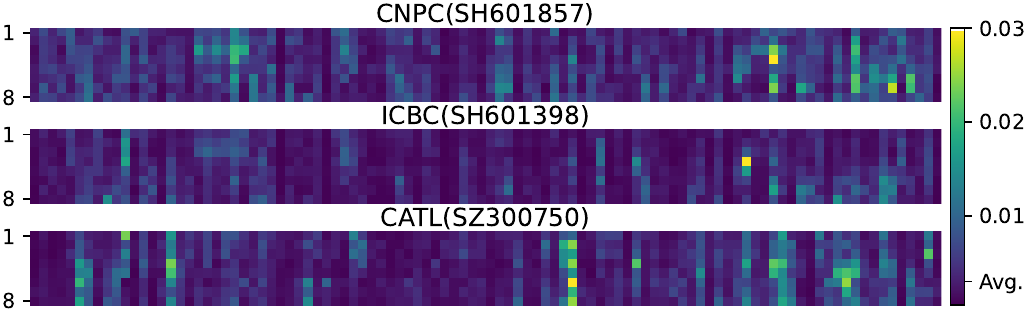}
\caption{The correlation towards three target stocks on Aug 19th, 2022. The y-axis is time steps in the lookback window and the x-axis is source stocks. \emph{Avg.} denotes the evenly distributed value.}
\label{fig:stock2date}
\end{figure}
\subsection{Stock Transformer Architecture (RQ2)}
We validate the effectiveness of our specialized stock transformer architecture by experiments on four settings. (1) (MA)STER, which is our stock transformer without the gating. (2) (MA)STER-Bi, in which we substitute the single-layer transformer encoder with a bi-directional LSTM to evince that the effectiveness of our proposed architecture is not coupled with strong sequential encoders.
(3) Naive, which directly performs information aggregation among $\tau \times |\mathcal{S}|$ tokens. 
(4) Clustering, in which we adapt the Local Sensitive Hashing~\cite{kitaev2020reformer} to allocate all tokens into $10$ buckets by similarity and perform aggregation within each group, which is a classic task-agnostic technique to reduce the scale of the attention field. 
For a fair comparison, in (3) and (4), we first use the same transformer encoder to extract token embedding and then use the same multi-head attention mechanism as in our stock transformer, so the only difference is the attention field.
Due to resource limits, we only conduct experiment on CSI300 dataset. The results in Table~\ref{tab:effect} illustrate the efficacy of our tailored stock transformer architecture, which performs intra-stock aggregation and inter-stock aggregation alternatively.

\subsection{Ablation Study (RQ3)}
First, we conduct ablation study on $(N_1,N_2)$ combination. 
The results of CSI300 are shown in Figure~\ref{fig:mh} and the results on CSI800 are similar. The difference among head combinations is not significant compared with the inherent variance under each setting. In the studied range, most settings consistently performed better than the baselines.

Second, we study the influence of temperature $\beta$ in the gating mechanism.
As explained before, a smaller $\beta$ forces a stronger feature selection while a larger $\beta$ turns off the gating effect.
Figure~\ref{fig:beta} shows the performance with varying $\beta$.
The CSI300 is a relatively easier dataset where most features are quite effective, so the temperature is expected to be larger to relax the feature selection, while more powerful feature selection intervention is needed for the sophisticated CSI800 dataset whose $\beta$ of the best performance is smaller.

\subsection{Visualization of Attention Maps (RQ4)}
\begin{figure}[t]
    \includegraphics[width=\linewidth]{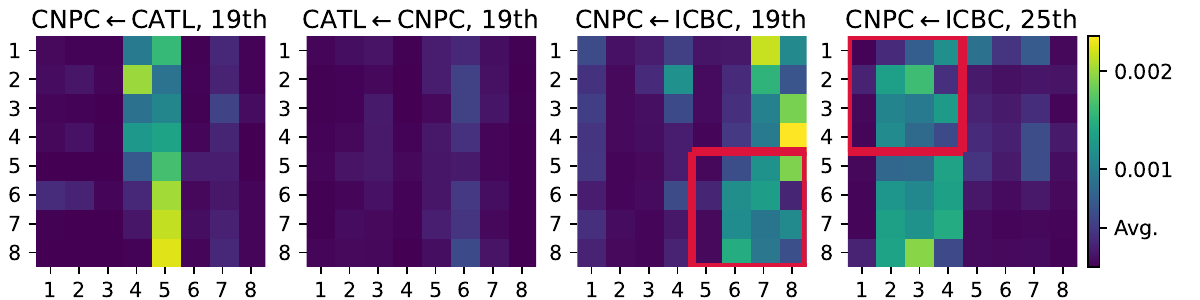}
\caption{Cross-time correlation of stock pairs on Aug 19th and 25th, 2022. The x-axis is the source time steps and the y-axis is the target time steps. }
\label{fig:stock2stock}
\end{figure}

We show how \frameworkname~captures the momentary and cross-time stock correlation that previous methods are not expressive enough to model.
Figure~\ref{fig:stock2date} shows the inter-stock attention map at different time steps in the lookback window. We choose three representative stocks as the target and sample $100$ random stocks as sources for visualization.
The highlighted part is scattered instead of exhibiting neat strips, implying that the correlation is momentary rather than long-standing.
Also, the inter-stock correlation is sparse, with only a few stocks having strong correlations toward the target stocks.
Figure~\ref{fig:stock2stock} displays the correlation between stock pairs to show how the correlation resides in time. 
From source stock $v$ to target stock $u$, we compute $I_{u\leftarrow v}[i,j]=\mathrm{S}^1_v[i,j]\mathrm{S}^2_i[u,v]$ as the $\tau\times\tau$ correlation map, while $\mathrm{S}^1$ and $\mathrm{S}^2$ are the intra-stock and inter-stock attention map.
First, the highlighted blocks are not centered on the diagonal, because the stock correlation is usually cross-time rather than temporally aligned. 
Second, the left two figures are totally different, illustrating that correlation is highly asymmetric between $u\leftarrow v$ and $v\leftarrow u$. 
Third, the importance of mined correlation changes slowly when the lookback window slides to forecast on different dates. For example, blocked regions in the right two figures correspond to the same absolute time scope of different prediction dates, whose patterns are to a certain degree similar. 

\section{Conclusion}
We introduce a novel method \frameworkname~for stock price forecasting, which models the realistic stock correlation and guide feature selection with market information.
\frameworkname~consists of five steps, market-guided gating, intra-stock aggregation, inter-stock aggregation, temporal aggregation, and prediction.
Experiments on the Chinese market with $2$ stock universe shows that \frameworkname~achieves averagely $13\%$ improvements on ranking metrics and $47\%$ on portfolio-based metrics compared with all baselines.
Visualization of attention maps reveals the de-facto momentary and cross-time stock correlation. 
In conclusion, we provide a more granular perspective for studying stock correlation, while also indicating an effective application of market information. Future work can explore to mine stock correlations of higher quality and study other uses of market information.

\section*{Acknowledgements}
The authors would like to thank the anonymous reviewers for their insightful reviews. 
This work is supported by the National Key Research and Development Program of China (2022YFE0200500), Shanghai Municipal Science and Technology Major Project (2021SHZDZX0102), and SJTU Global Strategic Partnership Fund (2021SJTU-HKUST).
\bibliography{13208.Li}
\end{document}